\documentclass[a4paper,fleqn,usenatbib]{mnras}

\usepackage[T1]{fontenc}
\usepackage{ae,aecompl}

\usepackage{graphicx}	
\usepackage{amsmath}	
\usepackage{amssymb}	
\usepackage{txfonts}
\usepackage{enumitem}
\usepackage{natbib}
\usepackage{xcolor}

\title[Impact of atmospheric dispersion]{The impact of atmospheric dispersion in the performance of high-resolution spectrographs}

\author[B. Wehbe et al.]{
B. Wehbe,$^{1,2}$\thanks{E-mail: bachar.wehbe@astro.up.pt}
A. Cabral,$^{3,4}$
J.H.C. Martins,$^{1}$
P. Figueira,$^{5,1}$ 
N.C. Santos,$^{1,2}$
and G.\'{A}vila,$^{6}$
\\
$^{1}$Instituto de Astrof\'{i}sica e Ci\^{e}ncias do Espa\c{c}o, Universidade do Porto, CAUP, Rua das Estrelas, 4150-762 Porto, Portugal\\
$^{2}$Departamento de F\'{i}sica e Astronomia, Faculdade de Ci\^{e}ncias, Universidade do Porto, Rua Campo Alegre, 4169-007 Porto, Portugal\\
$^{3}$Instituto de Astrof\'{i}sica e Ci\^{e}ncias do Espa\c{c}o, Universidade de Lisboa, Campus do Lumiar, Estrada do Pa\c{c}o do Lumiar 22, Edif. D, PT1649-038 Lisboa, Portugal\\
$^{4}$Departamento de F\'{i}sica, Faculdade de Ci\^{e}ncias, Universidade de Lisboa, Campo Grande 1749-016 Lisboa Portugal \\
$^{5}$European Southern Observatory, Alonso de C\'{o}rdova 3107, Vitacura, Regi\'{o}n Metropolitana, Chile \\
$^{6}$European Southern Observatory, Karl-Schwarzschild-Stra{\ss}e 2, 85748 Garching bei M\"{u}nchen, Germany
}

\date{Accepted XXX. Received YYY; in original form ZZZ}

\pubyear{2019}

\begin{document}
\label{firstpage}
\pagerange{\pageref{firstpage}--\pageref{lastpage}}
\maketitle

\begin{abstract}
	Differential atmospheric dispersion is a wavelength-dependent effect introduced by the atmosphere. It is one of the instrumental errors that can affect the position of the target as perceived on the sky and its flux distribution. This effect will affect the results of astronomical observations if not corrected by an atmospheric dispersion corrector (ADC). In high-resolution spectrographs, in order to reach a radial velocity (RV) precision of 10 cm/s, an ADC is expected to return residuals at only a few tens of milli-arcseconds (mas). In fact, current state-of-the-art spectrographs conservatively require this level of residuals, although no work has been done to quantify the impact of atmospheric dispersion. In this work we test the effect of atmospheric dispersion on astronomical observations in general, and in particular on RV precision degradation and flux losses. Our scientific objective was to quantify the amount of residuals needed to fulfil the requirements set on an ADC during the design phase. We found that up to a dispersion of 100 mas, the effect on the RV is negligible. However, on the flux losses, such a dispersion can create a loss of $\sim$2\% at 380 nm, a significant value when efficiency is critical. The requirements set on ADC residuals should take into consideration the atmospheric conditions where the ADC will function, and also all the aspects related with not only the RV precision requirements but also the guiding camera used, the tolerances on the flux loss, and the different melt data of the chosen glasses.
\end{abstract}

\begin{keywords}
atmospheric effects -- instrumentation: spectrographs -- methods: data analysis -- techniques: radial veocities
\end{keywords}



\section{Introduction}

Astronomical observations performed using ground-based telescopes are affected by wavelength-dependent atmospheric dispersion when taken at 
a zenithal angle different from zero. The atmospheric dispersion is due to the variation with wavelength of the refractive index of the atmosphere. 
By definition, an atmospheric dispersion model computes the refractive index of dry air at sea level as function of wavelength, n($\lambda$). 
Since observatories are usually located at higher altitudes, the refractive index should be corrected for temperature, pressure, and relative humidity (RH). 
From \cite{Smart1931}, we have that the atmospheric dispersion, in milli-arcseconds (mas) in the sky, is given by:
\begin{align}
   \Delta R (\lambda) &= R(\lambda) - R(\lambda_{\rm ref}) \nonumber \\
   \Delta R (\lambda) &\approx 206265 \left[n(\lambda) - n(\lambda_{\rm ref}) \right] \times \tan Z,
   \label{eq:dispersion}
\end{align}
where  $\lambda_{\rm ref}$ is the reference wavelength, and $Z$ is the zenithal angle of observation. \\
The problem of atmospheric dispersion is particularly serious for high-resolution spectrographs aiming at sub-pixel RV precision. In order to detect Earth-like planets orbiting Sun-like stars, a RV precision of around 10 cm s$^{-1}$ must be achieved. With the objective of developing high-resolution spectrographs that can reach such a RV precision \citep{Fischer2016}, several instrumentation challenges need to be solved including a correction of atmospheric dispersion variation down to this level of positioning accuracy. In order to reach the highest reproductibility in collected spectra and RV properties, an atmospheric dispersion corrector \citep[ADC,][]{Avila1997} is thus mandatory.
\\
Even assuming a perfect telescope centering and guiding at a given reference wavelength, atmospheric dispersion will shift the barycenter of the image as a function of wavelength \citep{Cohen1988}. This will strongly affect the light collected by a spectrograph, especially in high-resolution spectrographs aiming at high-fidelity spectra \citep{Blackman2019}. The loss of light, due to the elongated PSF that can be bigger than the size of the fiber, as a function of wavelength introduces a wavelength dependent photon-noise that becomes particularly stronger at shorter wavelengths where the dispersion is stronger, reducing the ability to measure efficiently the line's position and its properties. The importance of increasing the collected light at shorter wavelength comes from two aspects: 
\begin{enumerate}[label=(\roman*)]
\item the first from the fact that the efficiency of many optical components is lower at shorter wavelengths;
\item while the second is from the fact that around 380 nm, there's an abundance of CaII lines that are used as stellar activity indicators \citep{Santos2008}.
\end{enumerate}
In addition, imperfect atmospheric dispersion will change the continuum slope \citep{Pepe2008} which will alter the weight of spectral lines and influence the mean radial velocity (RV) value. \\

According to \cite{Blackman2019}, the ADC coupled with the injection on the fiber (controlled by a guiding camera) might affect the observations and introduce chromatic changes in the recorded spectra. To our knowledge, an exhaustive analysis of the problem in the visible range of the spectrum has never been performed. \cite{Bechter2018} estimated a RV error of 6 cm s$^{-1}$ for the combined adaptive optics (AO) and ADC correction in the near-infrared without mentioning the level of residuals. Also, \cite{Halverson2016} associated an error of 6.9 cm s$^{-1}$ from an ADC with 100 mas of residuals.
\\
While designing an ADC and setting the requirements on the residuals, one should pay attention to not only the effect of atmospheric dispersion residuals on RV, but also on the flux loss associated. The influence of optical glass melt data is also important. In fact, different batches from the manufacturer will end up with different Sellmeier coefficients \footnote{These are the coefficients of the Sellmeier equation that establish the relation between the refractive index and wavelength for a particular transparent medium; it is used to determine the dispersion of light in the medium.} due to melt procedures \citep{Langenbach1999}. The design of an ADC is based on catalog data that is not optimized to the melt data. From an instrumentation point of view, setting the requirements on the ADC to be so demanding (< 50 mas peak-to-valley (PV), maximum amount of residuals within the wavelength range of interest) is becoming more and more difficult. Not only the fact that finding glasses that can return such residuals, and the fact that atmopsheric models are not as accurate (to the level of the requirements), but also the manufacturing process is not as easy as it looks like. A manufacturing error of 1 arcsecond on the prisms angles will create a variation of 12\% on the residuals.\\
 In this paper, in order to fully understand the effects of ADC residuals, we will simulate its effect on synthetic spectra in section \ref{sec:simulations}, show the results and the analysis in sections \ref{sec:results} \& \ref{sec:discussion}.

 \section{Simulations}
 \label{sec:simulations}
 To test the effect of atmospheric dispersion on astronomical observations, and specially the impact on RV, we generated synthetic spectra affected by dispersion. These were created with a tool developed for the work of \cite{Martins2018} which create spectra that mimic reduced HARPS observations. HARPS, \citep[High-Accuracy Radial velocity Planet Searcher,][]{Mayor2003}, is one of the most precise spectrographs on Earth, with a design that was reproduced on HARPS-N and that greatly inspired ESPRESSO. It is considered the prototypical fiber-fed spectrograph. Since we had access to the HARPS pipeline, to expedite and simplify the reduction process with no loss of generality, we decided to use the HARPS spectrograph as our spectral template. It is important to state that no ADC is assumed in all the simulations during this work except for section \ref{subsec:adcresiduals}.
 
\subsection{Star and fiber simulations}
\label{subsec:star_simulations}
\begin{figure}
\centering
\includegraphics[width=\hsize]{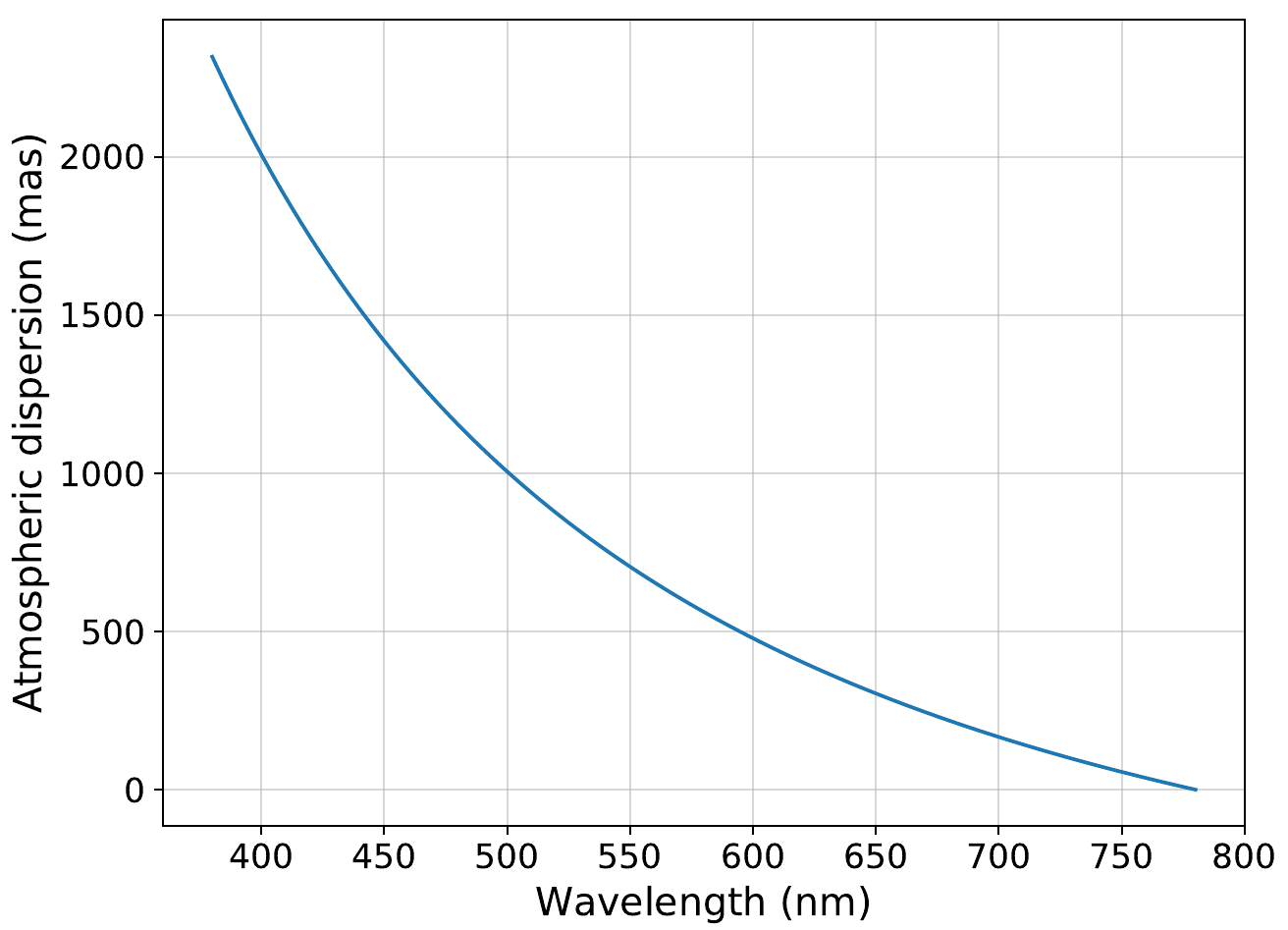}
\caption{Differential atmospheric dispersion for a zenith angle of 60$^{\circ}$ using the Filippenko model for La Silla observatory.}
\label{Fig:filippenko}
\end{figure}

\begin{figure}
\centering
\includegraphics[width=\hsize]{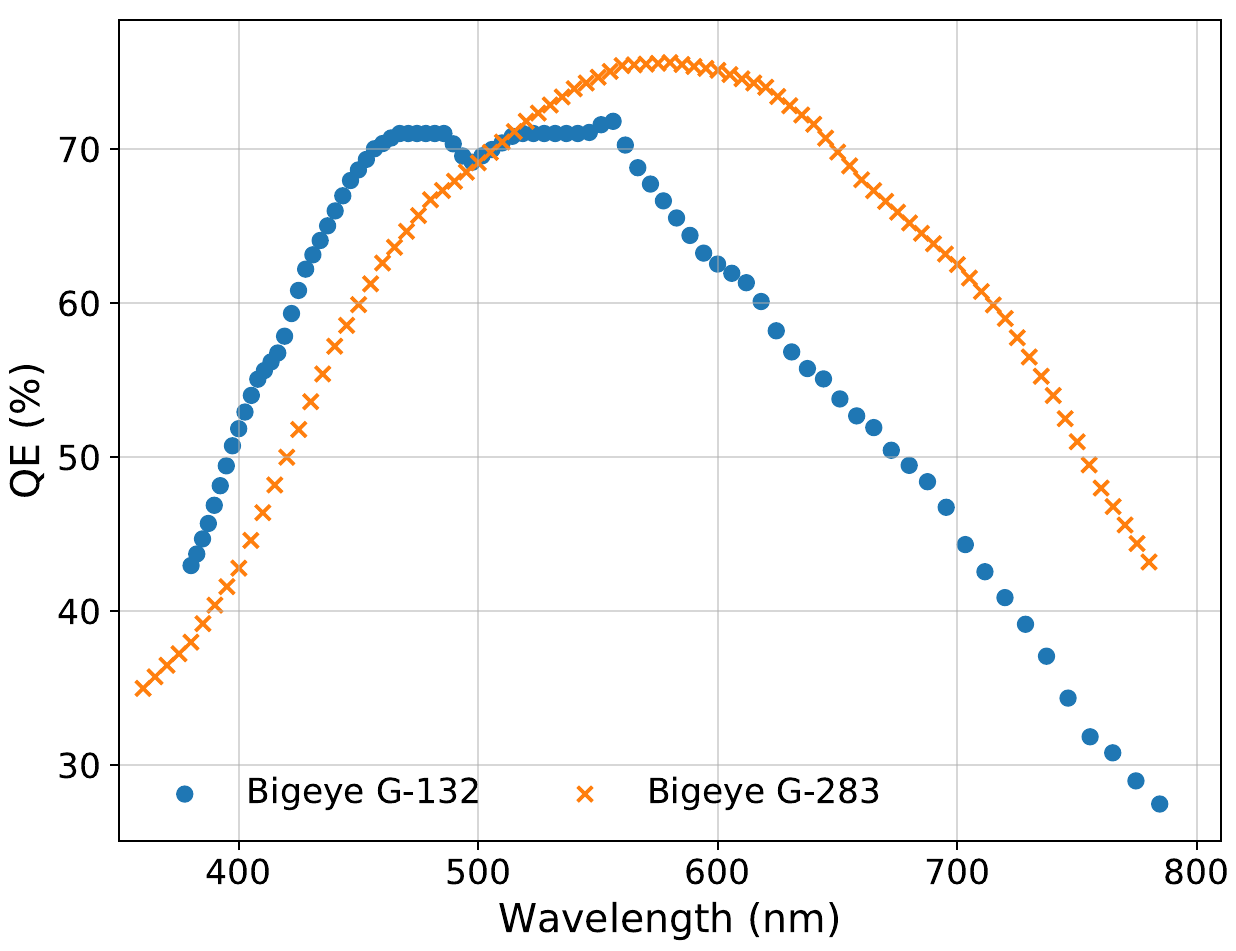}
\caption{Absolute quantum efficiency of Bigeye G-132 and G-283 scaled to wavelength range of interest.}
\label{Fig:qe}
\end{figure}

In order to simulate the point spread function (PSF) of the scientific object on the guiding camera, we start with a 2D Gaussian function: \\
$ I(x,y) = e^{-\frac{(x-x_0)^2}{2\sigma_x^2}-\frac{(y-y_0)^2}{2\sigma_y^2}}$ \\
where I is the intensity, (x,y) represent the displacement with respect to the origin ($x_0$,$y_0$); 
$\sigma_x$ and $\sigma_y$ represent the sigma of the distribution in arcseconds. Since we are interested in studying the effect of atmospheric dispersion, it is important to understand how it will affect the position of the barycenter of the PSF. The atmospheric dispersion is a function of wavelength which will shift the barycenter as function of wavelength as well. For a star at zenithal angle Z, the angular atmospheric dispersion is proportional to Z (see Eq. \ref{eq:dispersion}). At the focus of a telescope, this will produce a linear dispersion with a well defined direction \citep{Wynne1986}.
For simplification purposes but with no loss of generality, we will assume that this displacement is along the $x$-axis of the guiding camera, while the displacement along the $y$-axis will remain constant. We also assumed that the seeing is constant as a function of wavelength.
As for $\sigma_x$ and $\sigma_y$ they are functions of the seeing and will be considered equal to 0.8", a typical value of La Silla observatory where HARPS is installed. 
We will use the dispersion model presented by \cite{Filippenko1982}, hereafter the Filippenko model, to estimate the dispersion for the wavelength range of interest between 380 nm and 700 nm, the wavelength coverage of HARPS, 
(see Figure \ref{Fig:filippenko}). The atmospheric dispersion computed is always relative to a reference wavelength. In this case, we assumed that the reference wavelength is 380 nm, centered at the center of the guiding camera. The barycenter of the PSF corresponding to the reference wavelength is considered the origin of displacement ($x_0$,$y_0$). The fiber size is assumed to be 1", a value equal to the HARPS fiber aperture.
\\
In addition, we take into consideration in the simulations the quantum efficiency (QE) of the guiding camera by multiplying the intensity of each wavelength by the corresponding QE. The variation of QE with wavelength will affect the position of the barycenter of the PSF.
For that, we assumed two different cameras, one more efficient in the blue and the other being more efficient in the red, the Bigeye G-132 and G-283, respectively.
Figure \ref{Fig:qe} shows the variation of the QE of both cameras of interest where it is clear that the Bigeye G-132 has higher efficiency in the blue part of the spectra.
 
The atmospheric dispersion is also a function of the zenithal angle of observation (Z; see Eq. \ref{eq:dispersion}). Therefore we simulated a PSF, taking into consideration the amount
of dispersion (0 to 2 arcseconds) with Z taking the values between 0$^{\circ}$ and 60$^{\circ}$ presented in Table \ref{Table:disp}.
 
\begin{table}
\caption{Zenithal angles (and airmass) of interest and their corresponding levels of atmospheric dispersion for La Silla observatory.}             
\label{Table:disp}      
\centering                          
\begin{tabular}{c c c}        
\hline                
Z ($^{\circ}$) & Airmass & Dispersion (mas) \\    
\hline                       
0 & 1 &0 \\
5 & 1.004 & 115 \\
10 & 1.015 & 230 \\
15 & 1.035 & 350 \\
30 & 1.154 & 750 \\
45 & 1.414 & 1300 \\
60 & 2 & 2250 \\
\hline                                  
\end{tabular}
\end{table}
 
\begin{figure}
\centering
\includegraphics[width=\hsize]{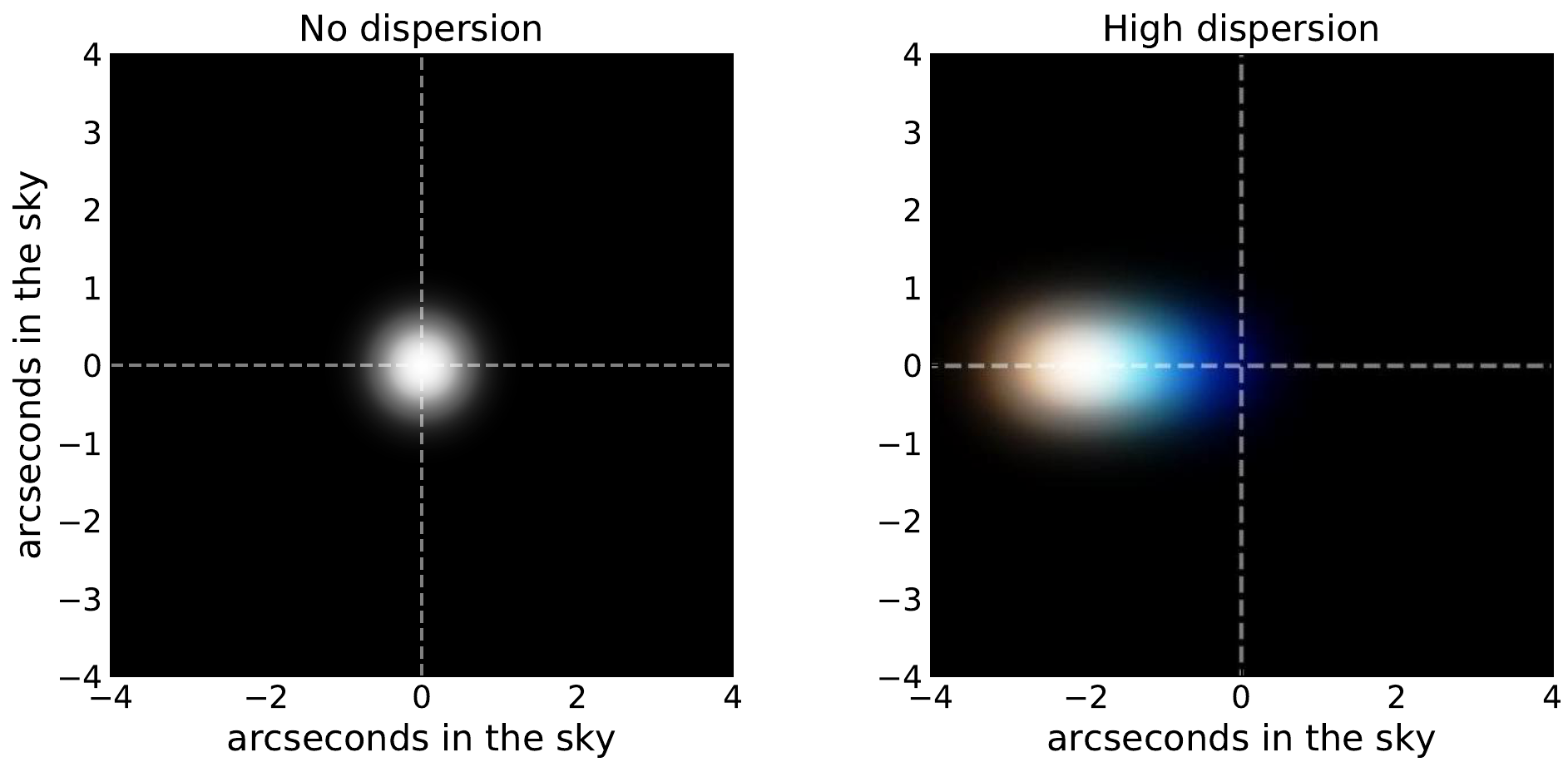}
\caption{Effect of atmospheric dispersion on the simulated spots taking into consideration the QE of the camera. Left: the case where there is no dispersion (Z = 0$^{\circ}$); right: the case where the dispersion is the highest (Z = 60$^{\circ}$). The dispersion is shown relative to the extreme blue wavelength. The dashed lines show the position of the origin (x$_0$,y$_0$).
}
\label{Fig:dispersion}
\end{figure}

\begin{figure}
\centering
\includegraphics[width=\hsize]{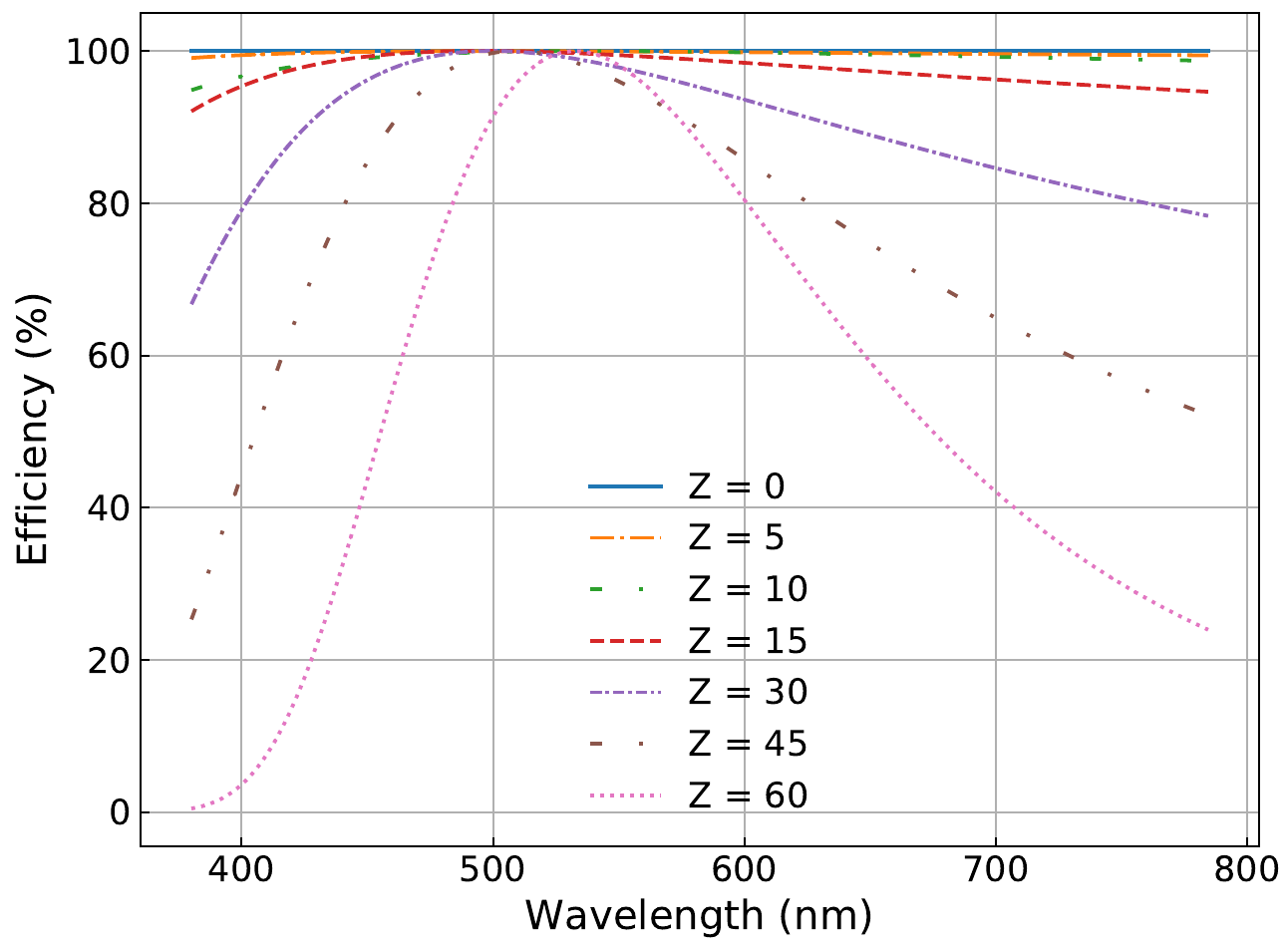}
\includegraphics[width=\hsize]{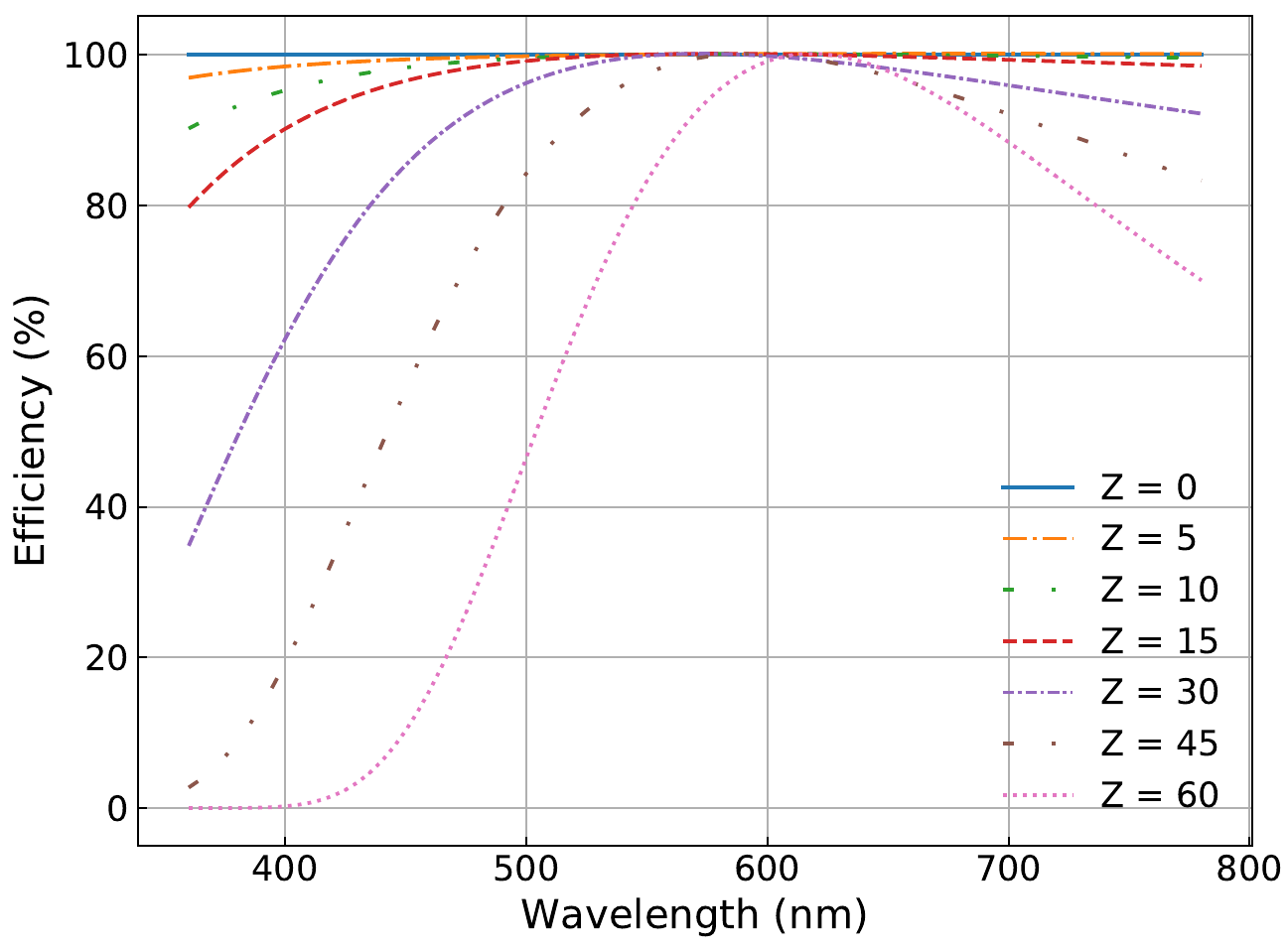}
\caption{Percentage of light of each wavelength entering the fiber for all values of Z. Bigeye G-132 (top); Bigeye G-283 (bottom).}
\label{Fig:fiber_eff}
\end{figure}

The case of no dispersion (Z = 0$^{\circ}$) will be used as reference in the remaining of this study. In Figure \ref{Fig:dispersion}, we show the effect of atmospheric dispersion on the shape of the PSF for 2 cases:
no dispersion (Z = 0$^{\circ}$), and high dispersion (Z = 60$^{\circ}$). The amount of dispersion will shift the position of the barycenter of the PSF; while the variation of QE will affect the intensity levels of each image. In the case of no dispersion, all the spots of different wavelengths will fall on the same location on the guiding camera and hence explaining the circular shape of the final spot. As for the case of high dispersion, each spot will be displaced from the reference spot (the spot corresponding to the reference wavelength) by an amount equal to the corresponding atmospheric dispersion which explain the elliptical shape of the final spot. 
\\
In order to compute the fraction of light at each wavelength that will enter the fiber, the fiber need to be centered at the barycenter of each spot for all the cases of interest, which will maximize the amount of light entering the fiber.
The barycenter is computed as follows:
\begin{equation}
\bar{x} = \dfrac{\sum\limits_{i=0}^{n}x_{i} \times m_{i}}{\sum\limits_{i=0}^{n}m_{i}} \\
\bar{y} = \dfrac{\sum\limits_{i=0}^{n}y_{i} \times m_{i}}{\sum\limits_{i=0}^{n}m_{i}} ,
\label{eq:barycenter}
\end{equation}  
where $m_i$ is the intensity of each point of coordinates $x_i,y_i$; and $n$ corresponds to the number of images used (equivalent to the wavelength step between 380 and 780 nm; 84 images in our case). Due to atmospheric dispersion, the elongation of the PSF will increase compared to the no dispersion case.
This will vary the fraction of each wavelength entering the fiber for different values of dispersion. 
From Figure \ref{Fig:fiber_eff}, we notice that with increasing Z, less light will enter the fiber and specially at the two limits of the wavelength range since the guiding cameras are usually wavelength centered (around 550 nm). We also notice that the position of the peak is not at the same wavelength with increasing Z. This variation indicates that the position of the barycenter is not constant. We can also notice that the amount of light entering, in the blue, is higher for the blue camera (Bigeye G-132), which is expected. In fact the guiding camera has a strong impact on which wavelengths are lost, and on which proportion. 
 
\subsection{Synthetic spectra simulations}
\label{subsec:spectra_simulations}
In order to simulate our synthetic spectra, we followed the method described by \cite{Martins2018}. It is important to state that in order to take into consideration the atmospheric dispersion effect on the spectra, the original code was updated to our requirements. During the design of an ADC, one of the important factors is the wavelength range where the ADC will be installed. Therefore, we were not interested in this work, in testing the effect of different spectral types and we limited our tests to the solar spectrum \citep{Kurucz2006} of G spectral type. Here we summarize the main steps and explain how the atmospheric dispersion effect
was taken into consideration:
\begin{itemize}
\item We started from the very high resolution (500000) solar spectrum of \cite{Kurucz2006}.
\item The spectrum  was degraded to the intended resolution of HARPS using the algorithm presented in \cite{Neal2017}.
\item The continuum was scaled in flux to match the expected flux of the target as computed from HARPS exposure time calculator (Table \ref{Table:ETC}).
\begin{table}
\caption{HARPS ETC settings. The settings not in this table were left to their default values.}             
\label{Table:ETC}      
\centering                          
\begin{tabular}{l r}        
\hline                
\textbf{Target input flux} &  \\    
\textbf{distribution} & \\
Template spectrum & G2V (Kurucz) \\
\textbf{Sky conditions} & \\
Airmass & Table \ref{Table:disp} \\
Seeing & 0.8" \\
\textbf{Instrument set-up} & \\
Exposure time & 300s, 100s, 50s \\
Signal to Noise & 220, 130, 90 \\ 
\hline                                  
\end{tabular}
\end{table}
\item Each spectrum is then multiplied by the corresponding fiber efficiency ( Figure \ref{Fig:fiber_eff}) based on the assumed zenithal angle (amount of dispersion).
\item The noise is assumed to be Gaussian photon noise.
\end{itemize}
We simulated 10$^4$ spectra with different photon noise component for each dispersion case.
These spectra will be then used to test and evaluate the effect of atmospheric dispersion.
 
 \subsection{From synthetic spectra to radial velocities}
 \label{subsec:spectrarv}
 In order to reduce these spectra and compute the RV, HARPS pipeline (version 3.4) was used.
 We produced two sets of reduced data by turning on and off the flux correction function, which is a pipeline variable.
 The flux correction function will compare the flux from each spectra to a template based on the spectral type of the target.
 These templates are built from very high signal-to-noise ratio (SNR) observations \citep{espr2019} of standard stars at low airmass. 
 Then, it will apply a correction in order to scale the flux to the same level of the template by multiplying it by the ratio of flux between the observed and the template.
 In that case, all our spectra will have the same flux distribution as the template, which ensures that variable atmospheric conditions will not 
 induce any systematic effects in the cross-correlation function (CCF) computation. In case the correction was too small or too big due to mismatch with the flux template,
 the flux correction function is automatically switched off. In fact, according to the pipeline, if the ratio between the observed flux and the template is lower than 0.25 or higher than 3, the function will be off. We will not focus here on these limits as it is behind the scope of this paper. \\ In our case, the spectra were simulated assuming a G2 solar spectral type. 
 Based on HARPS pipeline, no flux template exist for this spectral type. Hence, we built our own G2 flux template using the same technique described above. We simulated a very high SNR spectrum at an airmass of 1, where the dispersion is null. In this study, all the spectra were corrected using our template. As mentioned above, HARPS pipeline was used to compute the RV. The RV in HARPS pipeline is computed following the CCF technique where high-resolution spectra are cross-correlated with numerical masks over a range around the host's RV \citep{Baranne1996}. The CCF function from the HARPS pipeline is used with the same inputs for all the spectra tested of all the cases. In that case we guarantee that the variation of RV, if any, is only due to the atmospheric dispersion effect.

\section{Results}
\label{sec:results}
As mentioned before, several tests were done in order to understand and evaluate the effect of the atmospheric dispersion to be able 
to justify the requirements on the residuals of an ADC.

\begin{figure}
\centering
\includegraphics[width=\hsize]{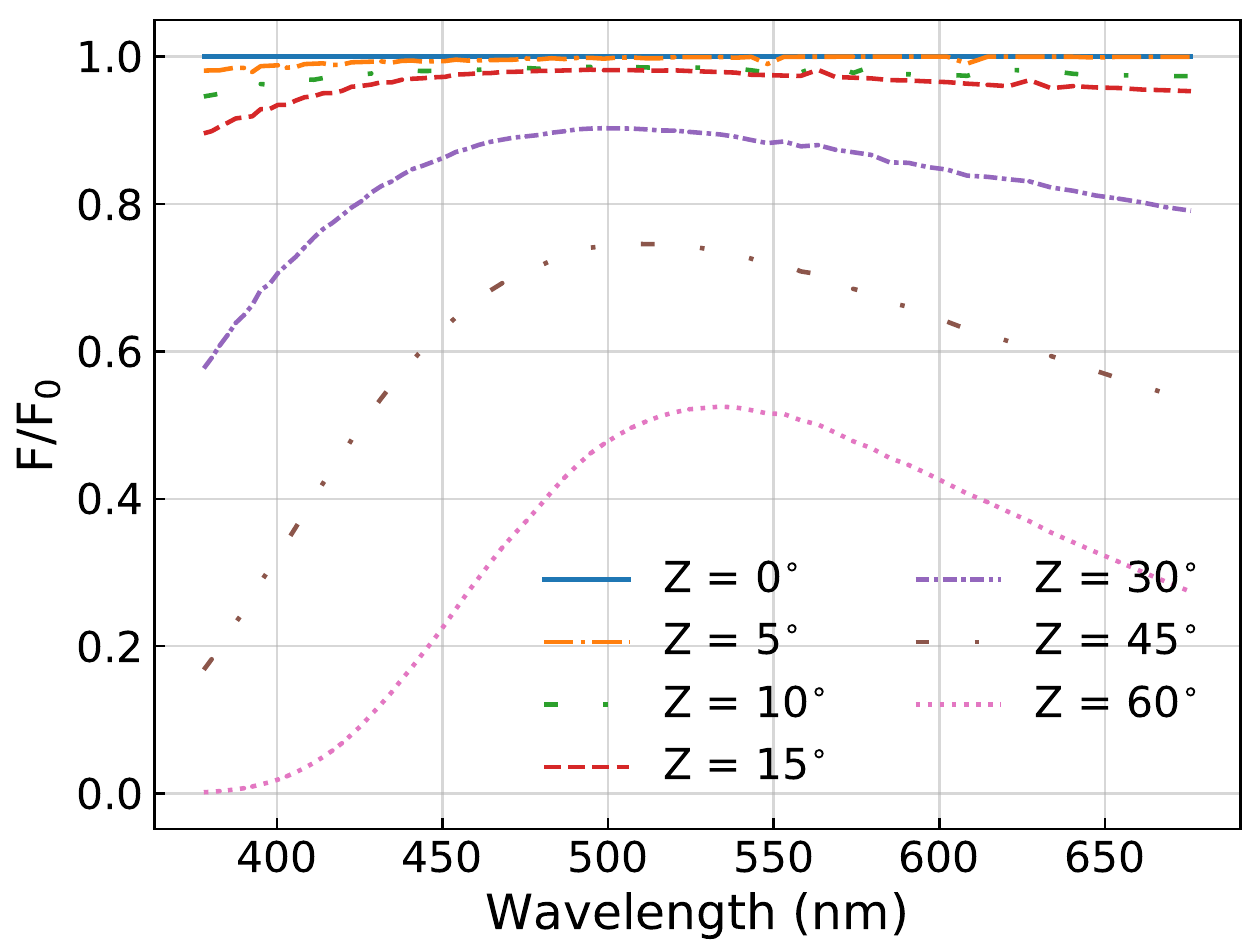}
\caption{Continuum slope variation for different Z cases compared to the reference Z = 0$^{\circ}$. The slopes were extracted from the synthetic spectra affected by atmospheric dispersion not corrected by an ADC.}
\label{Fig:slope}
\end{figure}

As well known, and discussed, e.g. in \cite{Pepe2008} and \cite{Fischer2016}, atmospheric dispersion will introduce a slope variation in the spectral continuum.
To verify this effect, we divided spectra, simulated with atmospheric dispersion not corrected by an ADC, of different zenithal angles by the reference (Z = 0$^{\circ}$) to see if there is any slope variation.
This effect can be clearly seen in Figure \ref{Fig:slope}. We only show four cases for visualization purposes. It is clear that starting at Z = 15$^{\circ}$, the slope variation starts to be more severe until it reaches the worst case for Z = 60$^{\circ}$ where a big part of the blue light is lost due to atmospheric dispersion. This slope variation should be corrected with the flux correction function at the level of the CCF. 

\subsection{RV precision degradation}
\begin{figure}
   \centering
   \includegraphics[width=\hsize]{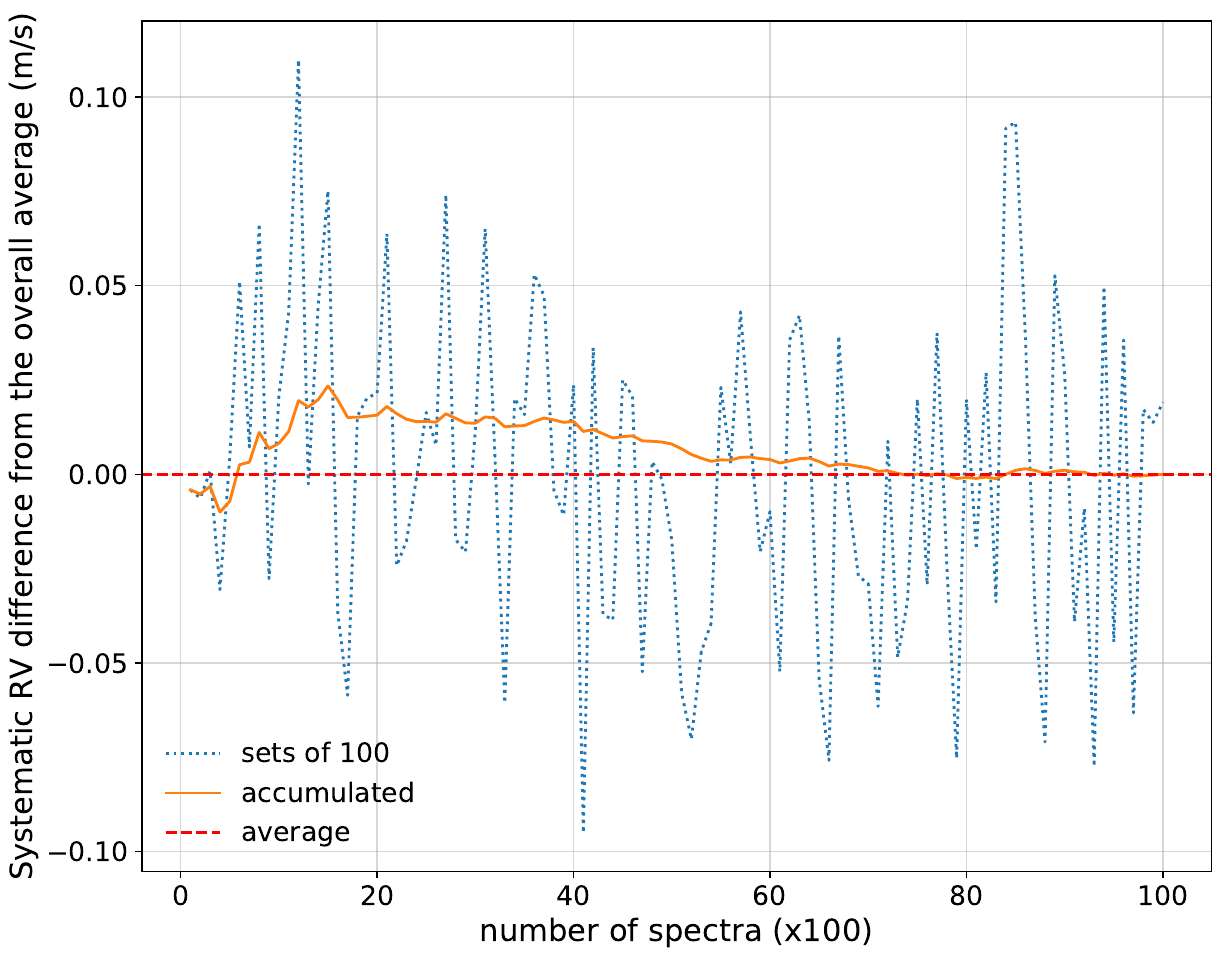}
   \caption{Systematic RV variation as function of number of simulations. It is clear that after 7000 spectra the results converge to the average value plotted as red dashed line.}
   \label{Fig:added}
\end{figure}

In order to test the effect of atmospheric dispersion on RV precision, if any, we simulated several spectra for each zenithal angle in Table \ref{Table:ETC}. To be able to characterize the effect of both random noise and systematic effects, 10$^4$ spectra were generated for each case assuming no ADC to correct the atmospheric dispersion. From Figure \ref{Fig:added}, we notice that after 7000 spectra, the results converge to the average value, allowing us to obtain both random and systematic error information. We also tested that the spectra have a gaussian distribution, well characterized by its mean and standard deviation.

\begin{figure}
\centering
\includegraphics[width=\hsize]{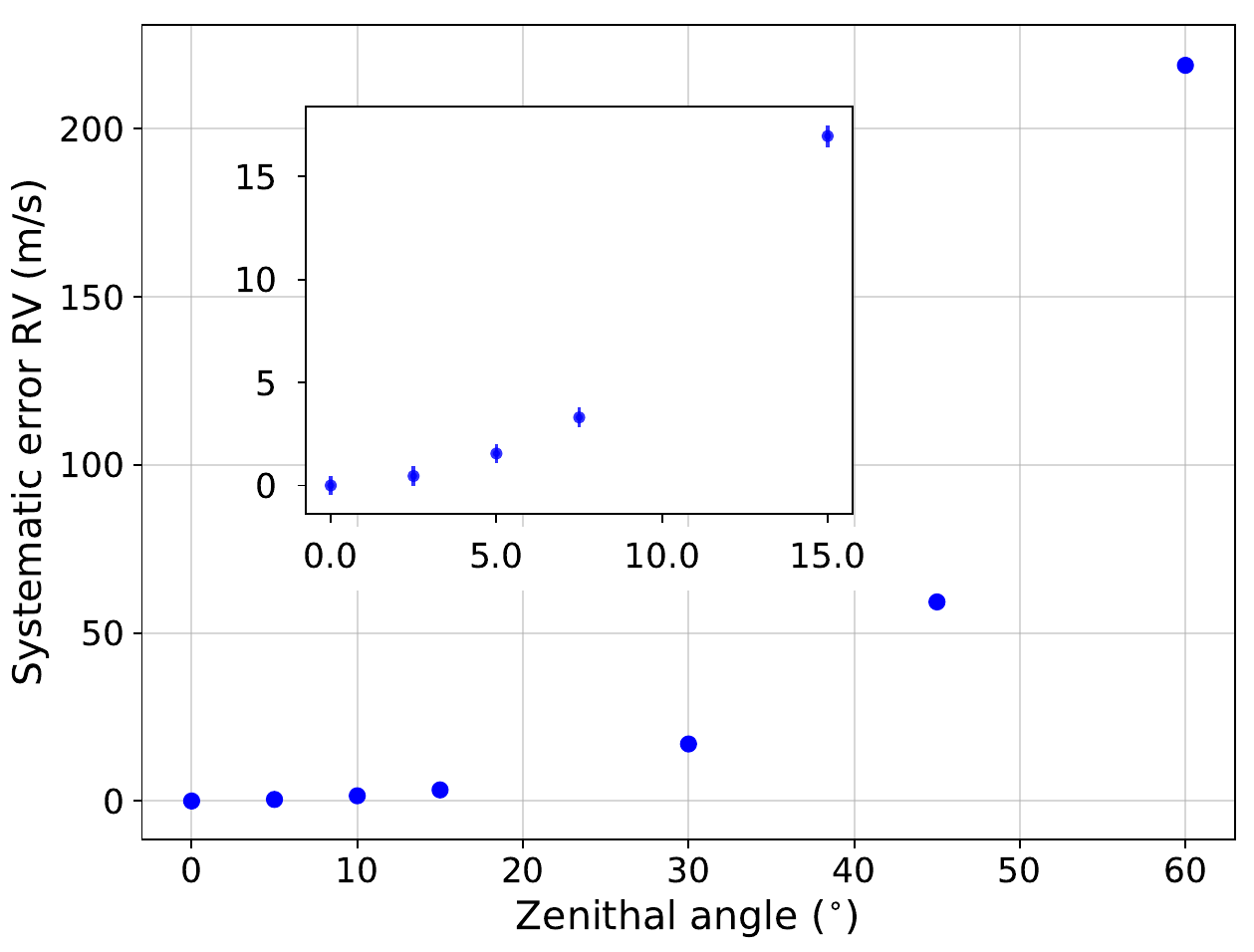}
\caption{Computed RV values of each set of simulated spectra when the flux correction function was \textbf{off}. The error bars correspond to the 1 $\sigma$ random error.}
\label{Fig:rv_off}
\end{figure}

\begin{figure}
\centering
\includegraphics[width=\hsize]{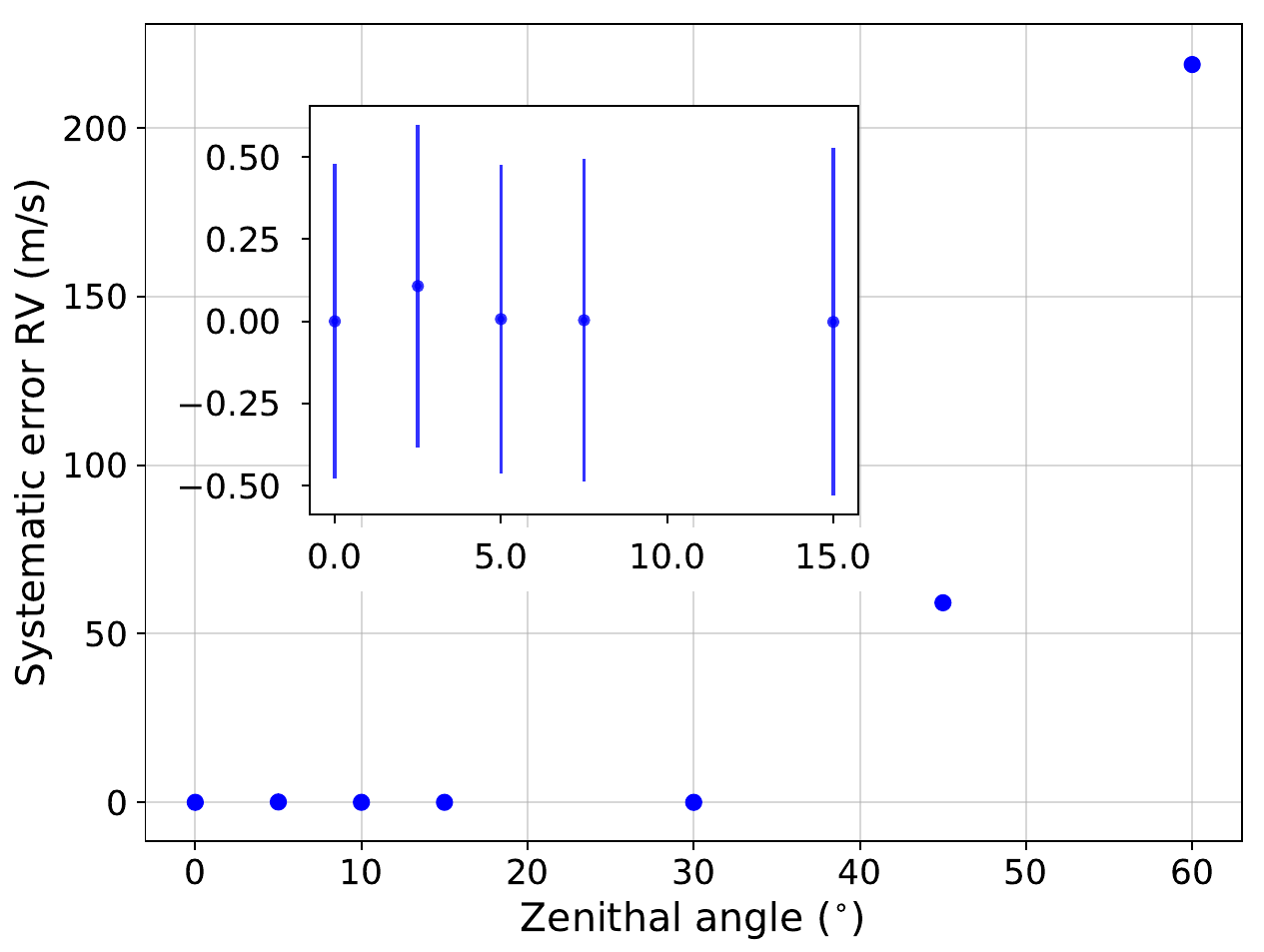}
\caption{Computed RV values of each set of simulated spectra when the flux correction function was \textbf{on}. The error bars correspond to the 1 $\sigma$ random error.}
\label{Fig:rv_on}
\end{figure} 

As it was shown before (Figure \ref{Fig:slope}), the atmospheric dispersion will introduce a slope variation of the continuum. This variation will hence give different weights 
to the spectral lines that will be used in the computation of the CCF. This will introduce a variation in the final RV. In figure \ref{Fig:rv_off}, we plot 
this variation in RV as function of zenithal angles. It is clear that this variation increases with Z, and hence with the amount of increasing dispersion. 
By definition, the flux correction function is supposed to correct for these atmospheric conditions variation which is also clear in Figure \ref{Fig:rv_on}. For a zenithal angle above Z = 40$^{\circ}$, the flux correction function is automatically switched off in the pipeline, explaining the sudden increase in the systematic error. The error bars shown are the random errors due to photon noise. \\
Since we are interested only in the systematic variation of the RV, which corresponds to the RV variation of all the zenithal cases compared to the one where no atmospheric dispersion was considered (zero-point RV at Z = 0$^{\circ}$), we decided to turn off the noise component in the simulated spectra in the rest of the paper, as we believe that the photon noise might affect our analysis. In this case, we will be only testing the effect of atmospheric dispersion.

\subsection{Flux losses due to different guiding cameras}
Due to the fact that atmospheric dispersion will increase the elongation of the PSF of the target on the focal plane where light is injected in the fiber (and imaged by the guiding camera), flux losses might be introduced. In fact, the atmospheric dispersion, if not corrected by an ADC, can be higher than 2", a value that is twice the typical size of the fibers used to inject the light into a spectrograph. Another problem, is the issue of guiding. Different guiding cameras, can lead to different fractions of each wavelength entering the fiber due to different QE and guiding algorithms. In this section, we will only focus on the effect due to different QE. We tested two different cameras in order to understand the effect of atmospheric dispersion on the variation of the barycenter position, the amount of flux that will enter the fiber, and the effect on the RV, if any.
\\
In order to test the effect of atmospheric dispersion on the position of the barycenter, we simulated the spot of the target as described in section 2.
These simulations were done for 2 different CCDs: Bigeye G-132 and Bigeye G-283, with high QE in the blue part of the spectra and the red one, respectively. \\

\begin{figure}
\centering
\includegraphics[width=\hsize]{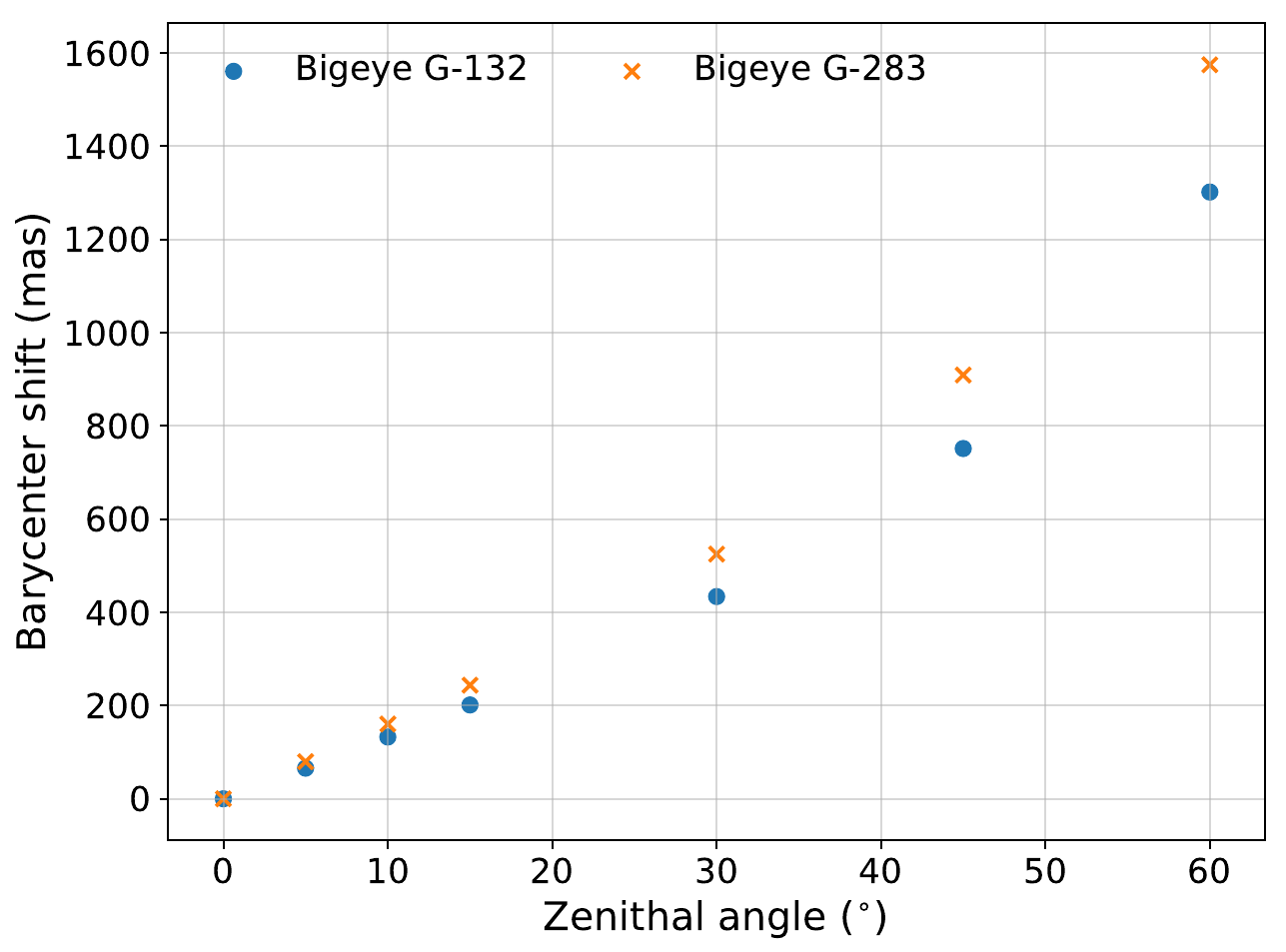}
\caption{Variation of the position of the barycenter with the variation of zenithal angle, for the two cameras of interest. These positions were measured relative to the position of no atmospheric dispersion (Z = 0$^{\circ}$) as reference.}
\label{Fig:bary_disp}
\end{figure}

We were able to measure the position of the barycenter on each case of Z using Eq. \ref{eq:barycenter}.
These positions were plotted as function of atmospheric dispersion (Figure \ref{Fig:bary_disp}). As expected, the position of the barycenter is proportional to the amount of dispersion: the higher it is, the further the barycenter is from the origin.

\begin{figure}
\centering
\includegraphics[width=\hsize]{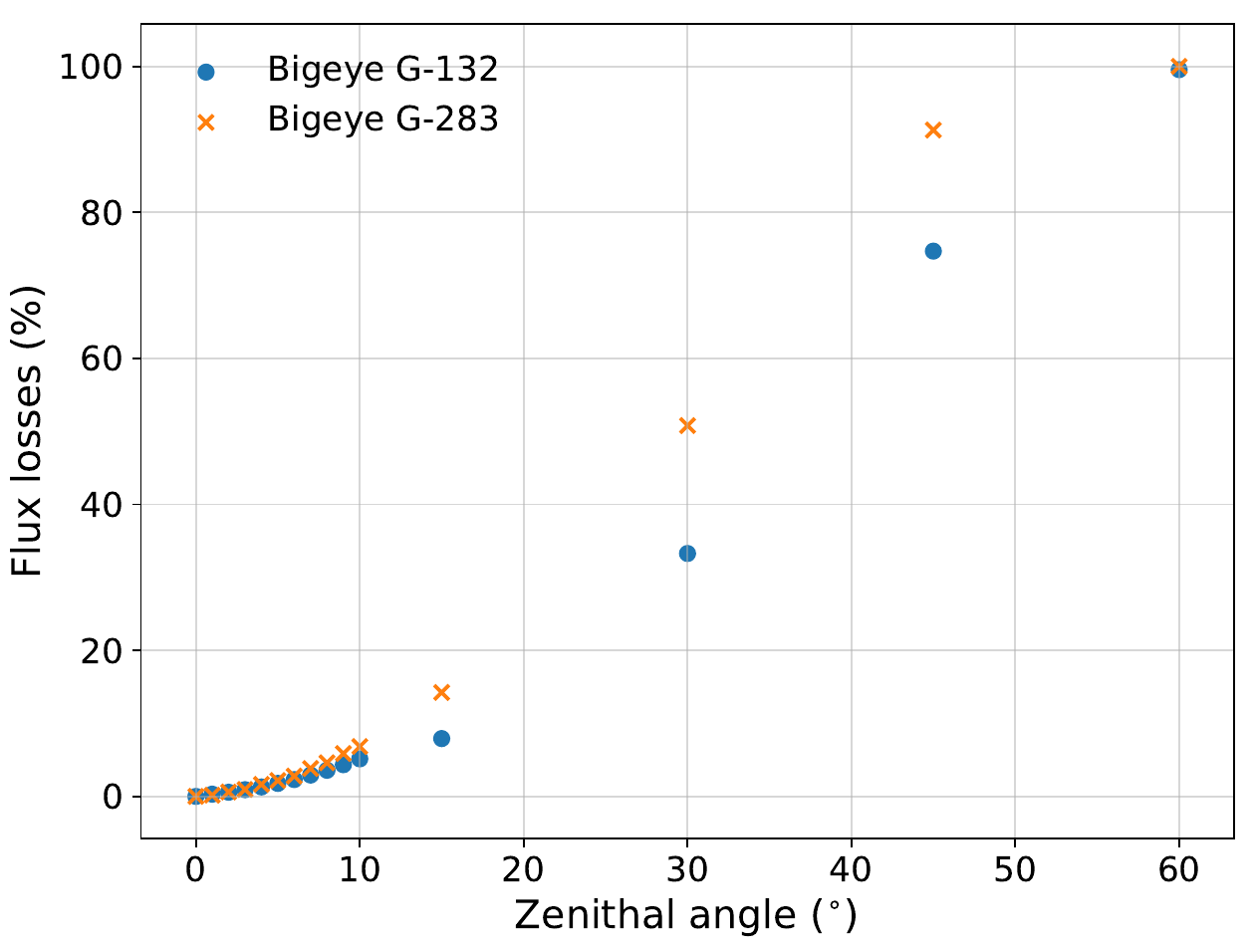}
\caption{Maximum flux loss at 380 nm for the two cameras of interest.}
\label{Fig:flux_loss_max}
\end{figure}

We also computed the maximum flux loss at 380 nm for the two cameras tested. It is clear from Figure \ref{Fig:flux_loss_max} that the choice of the guiding camera can increase or decrease the flux losses. In Figure \ref{Fig:flux_loss_max}, for Z = 60$^{\circ}$, the 2 cameras of interest suffers from total loss of flux at 380 nm due to the huge amount of atmospheric dispersion.

\subsection{Flux losses due to different melt data}
\label{subsec:melt}
\begin{figure}
   \centering
   \includegraphics[width=\hsize]{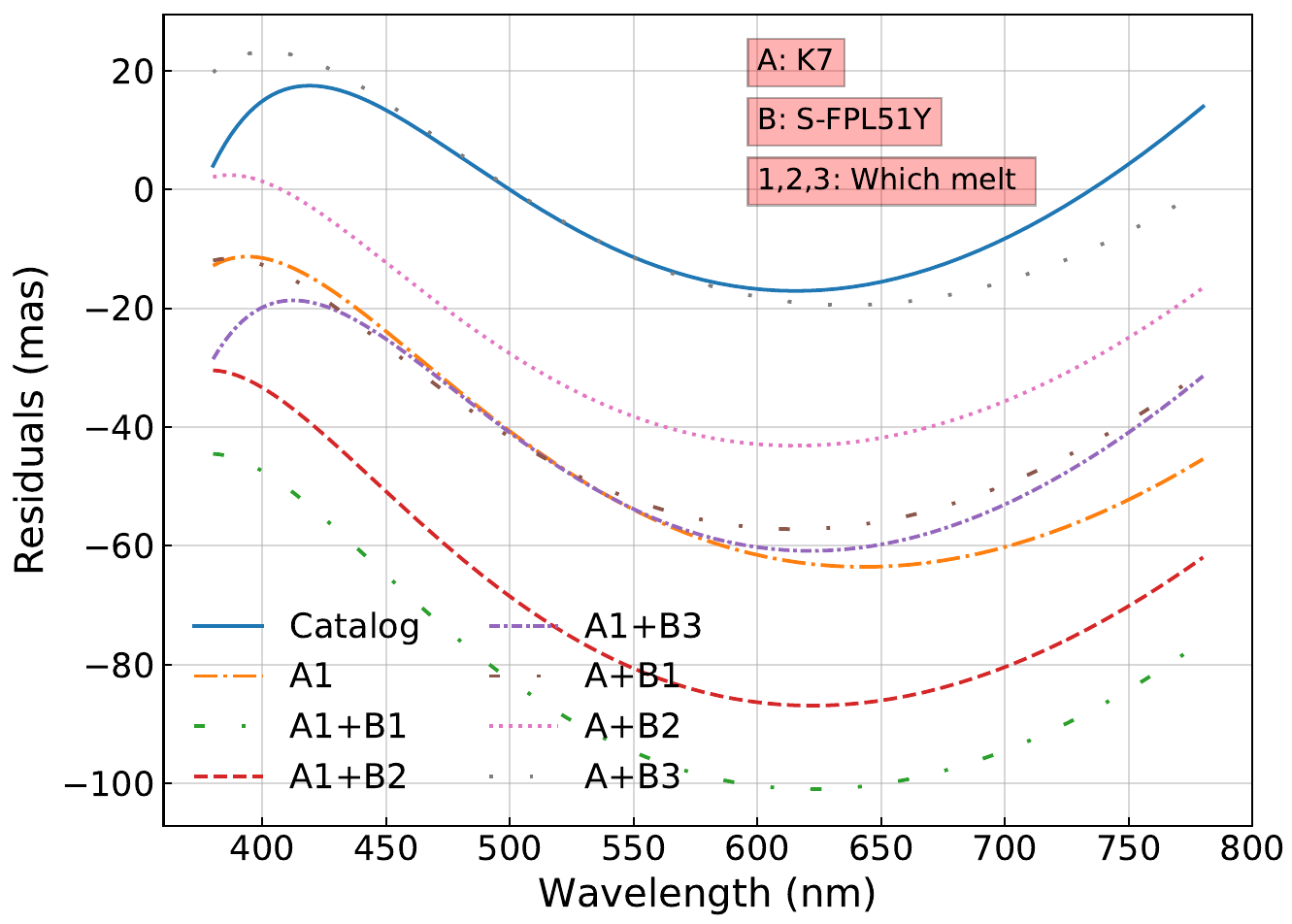}
   \caption{Residuals for all the melt data cases of  K7 + S-FPL51Y.}
   \label{Fig:melt_data_residuals}
\end{figure}

The melt data, glass characteristics of each batch produced, can cause a serious problem when designing an ADC \citep{Wehbe2019}. In fact, different melt data result with different Sellmeier coefficients
than the ones in the catalog normally used during the design phase of an ADC. \cite{Wehbe2019} provides a detailed analysis of the different cases that we will not 
mention here. We will use one of the cases, to test the effect, on the flux losses. In fact, different cases of melt data will result different shapes of 
ADC residuals and different PV that might alter the flux losses specially in the blue wavelengths. The variation between the melt cases will result in a variation in the Sellmeier coefficients that will affect the computed index of refraction. The latter will introduce a variation in the dispersion of the medium and hence the amount of residuals expected from the ADC. In Figure \ref{Fig:melt_data_residuals}, we show the different shapes and amounts of residuals
of different melt data (Table \ref{Table:melt}) for an ADC combination of K7 + S-FPL51Y \citep[the glass combination used for the ADC of ESPRESSO,][]{Cabral2012}. It is also important to state that the only variable between these curves was the melt data. The 
prism angles and the reference wavelength were maintained as computed during the design phase using the catalog Sellmeier coefficients. It is clear that not only the shape and the amount is changing in Figure \ref{Fig:melt_data_residuals}, but also the value of the residuals for the wavelength of 380 nm. In the catalog case, the ADC was designed to have a minimum of residuals at 380 nm. Different melt data, return different residuals at 380 nm. This variation is severe when it comes to flux losses.

\begin{figure}
   \centering
   \includegraphics[width=\hsize]{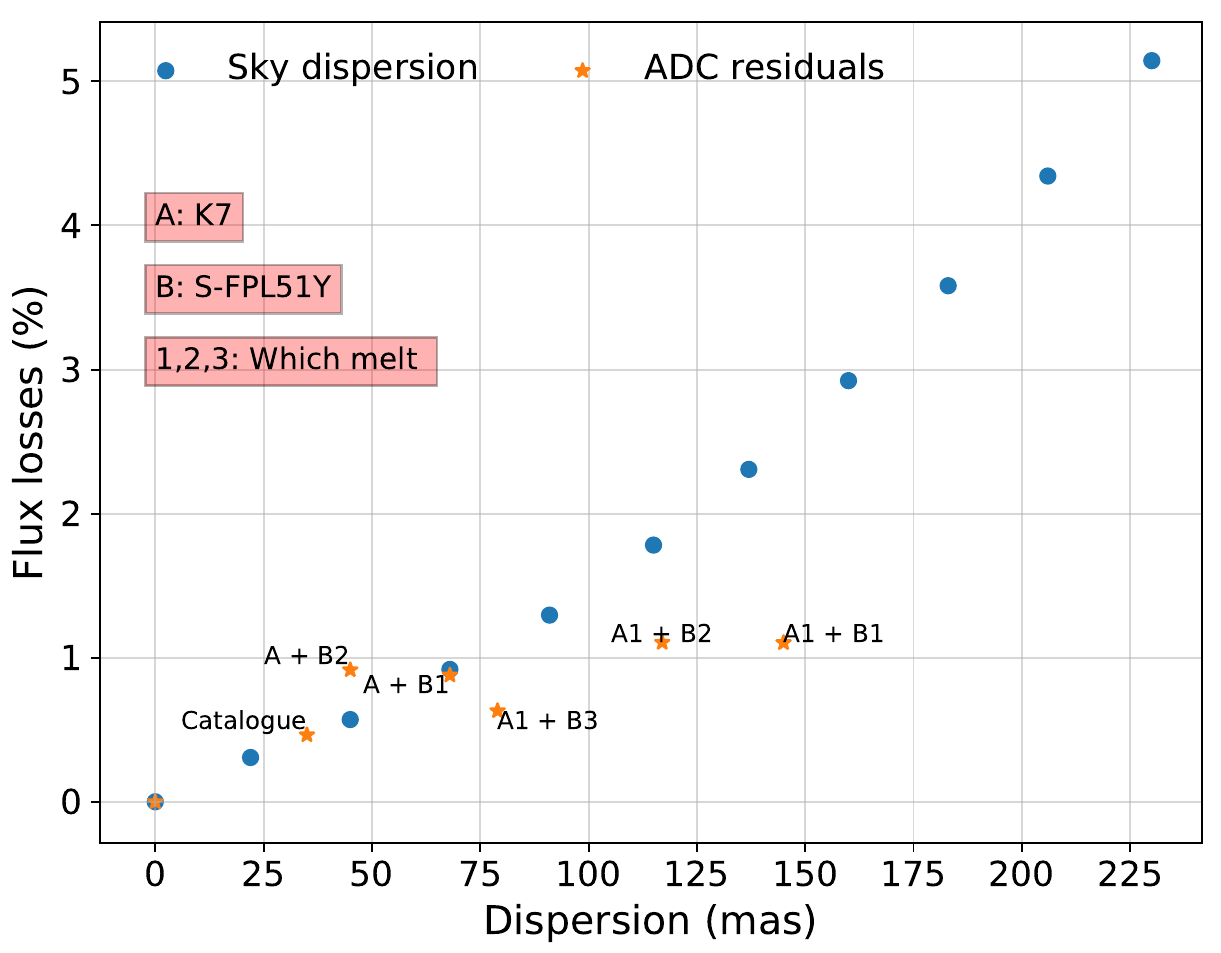}
   \caption{Maximum flux loss at 380 nm for all the melt cases.}
   \label{Fig:melt_data_flux_loss}
\end{figure}

In order to test the effect of melt data on flux losses, each residuals curve from Figure \ref{Fig:melt_data_residuals} was used to create a fiber efficiency injection curve as described in section \ref{subsec:star_simulations}.
These new fiber curves were then used to compute the flux floss at 380 nm for each case. The results are shown in Figure \ref{Fig:melt_data_flux_loss}. It is clear that different melt data cases, with different PV, similar to sky dispersion values, return different flux losses. This variation between the sky dispersion and the ADC residuals is due to the variation of the shape of the dispersion between on-sky and ADC.

\begin{table}
   \caption{Melt data designation}             
   \label{Table:melt}      
   \centering                          
   \begin{tabular}{l r}        
   \hline                
   A & K7 \\
   B & S-FPL51Y \\
   No number & Catalogue data \\
   With number & Melt data  \\
   \hline                                  
   \end{tabular}
   \end{table}

\section{Discussion}
\label{sec:discussion}
A key role when designing an ADC, is the amount of residuals set by the requirements (50 mas in the case of ESPRESSO). The tougher the requirements are, the harder it is to design, as there are less options in terms of glass pairs and the higher will be the cost. Building an ADC capable of returning the required residuals is not only a manufacturing challenge, but also a design challenge from chosing the correct glasses, to the correct angles, to the choice of the reference wavelength where the ADC will be optimized and the choice of the guiding camera and its QE. In fact, QE effect can be corrected with the use of a spectral filter in the guiding camera, nevertheless this is an option that should be analysed carefully as it would reduce the number of photons on the guiding camera and could jeopardize the efficiency of this task that has a huge impact on the total observation flux efficiency. All these factors, play a role when setting the residuals amount that, to our knowledge, is not a straightforward and well investigated area. In order to be able to set the requirements, we need to start by evaluating the effects of atmospheric dispersion on astronomical observations to be able to define the residuals of an ADC. In this section we will do so. 

\subsection{Impact of atmospheric dispersion}
\begin{figure}
   \centering
   \includegraphics[width=\hsize]{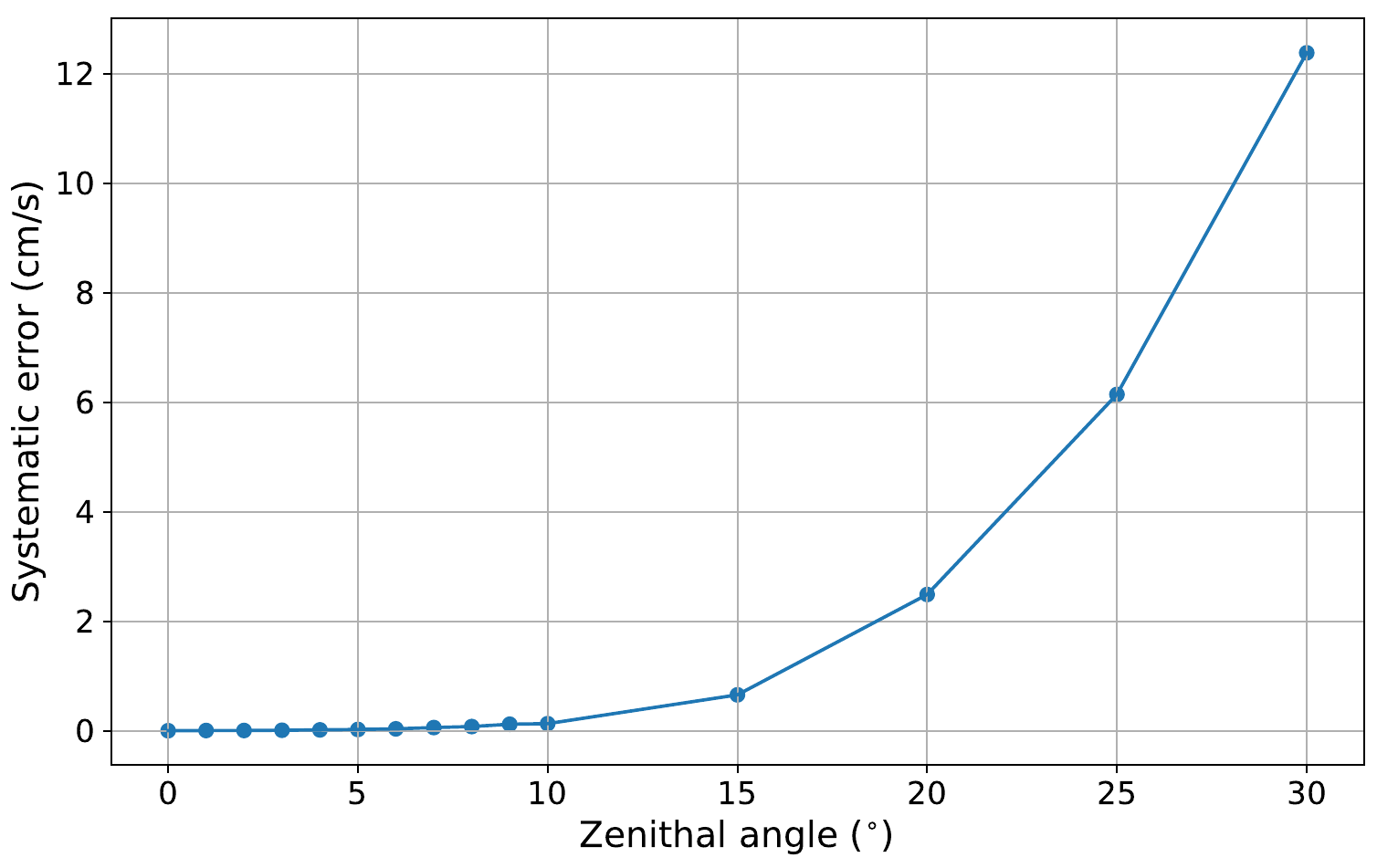}
   \caption{Systematic error variation as function of zenithal angle. The systematic error is behaving as expected, increasing with increasing Z.}
   \label{Fig:systematic_nonoise}
\end{figure}

Without an ADC, the atmospheric dispersion can be easily modelled using the Filippenko's model taking into consideration the different atmospheric parameters (temperature, pressure, RH and Z). By varying the zenithal angle, one can vary the amount of atmospheric dispersion as the shape will remain constant and similar to Figure \ref{Fig:filippenko}. It is clear from Figure \ref{Fig:rv_off} that the systematic error on RV start to be larger than 1 m/s for Z = 5$^{\circ}$ when the flux correction function is off. \\
When the flux correction function is on, these systematic errors are reduced to below 0.2 cm/s for Z = 5$^{\circ}$ (see Figure \ref{Fig:systematic_nonoise}) which implies the need of such a function in the pipelines. It also implies that an ADC is certainly needed to reduce the systematic errors to the level of 0.1 m/s. \\
To go further into investigation, we simulated spectra using the same technique described in \ref{subsec:spectra_simulations}, but this time by setting the noise to zero. In that case we will be able to fully characterize the systematic errors due to atmopsheric dispersion. It is clear from Figure \ref{Fig:systematic_nonoise}, that the effects of atmospheric dispersion up to Z = 5$^{\circ}$ (equivalent to $\sim$ 100 mas), on RV, are almost negligible (for a 0.1 m/s objective). With increasing zenithal angle, hence increasing the amount of dispersion, the amount of blue light of the spectra that enters the fiber decreases (see Figure \ref{Fig:fiber_eff}). The loss of light, lines in the spectra, will increase the error on RV as seen in Figure \ref{Fig:systematic_nonoise} with increasing Z.

\subsection{Impact of ADC residuals}
\label{subsec:adcresiduals}
Since an ADC is crutial in order to reduce the effects of atmospheric dispersion on astronomical observations, it is important to understant the effect of the ADC residuals. This effect, in general, should be similar to the effect of atmospheric dispersion by decreasing the zenithal angle to 5$^{\circ}$ or below which is corresponding to a dispersion of 100 mas and below, a value close to the current ADC residuals. In fact, the shape of the ADC residuals is not similar to the shape of the atmospheric dispersion, this is why we decided to use the ADC residuals curve (we simulated the case of ESPRESSO's ADC \citep{Cabral2012}, Figure \ref{Fig:melt_data_residuals}) instead of the Filippenko's curve (Figure \ref{Fig:filippenko}) and do a similar test as it was done previously in this work. It is also important to state the fact that ADC residuals are also proportional to Z. In this subsection, we will consider only the residuals corresponding to Z = 60$^{\circ}$, which is the worst case scenario. The residuals curves shown in Figure \ref{Fig:melt_data_residuals}, are used to produce curves similar to Figure \ref{Fig:fiber_eff}, which are used then to simulate spectra assuming the existence of an ADC. These new spectra were used to compute the RV as described in section \ref{subsec:spectrarv}. In this subsection, we will evaluate the effect of atmospheric dispersion residuals in mas instead of zenithal angles. \\
As discussed in subsection \ref{subsec:melt}, different melt data will return different amounts and shapes of ADC residuals. We will compare the different cases of melt, with the atmospheric dispersion results of the same amounts.

\begin{figure}
   \centering
   \includegraphics[width=\hsize]{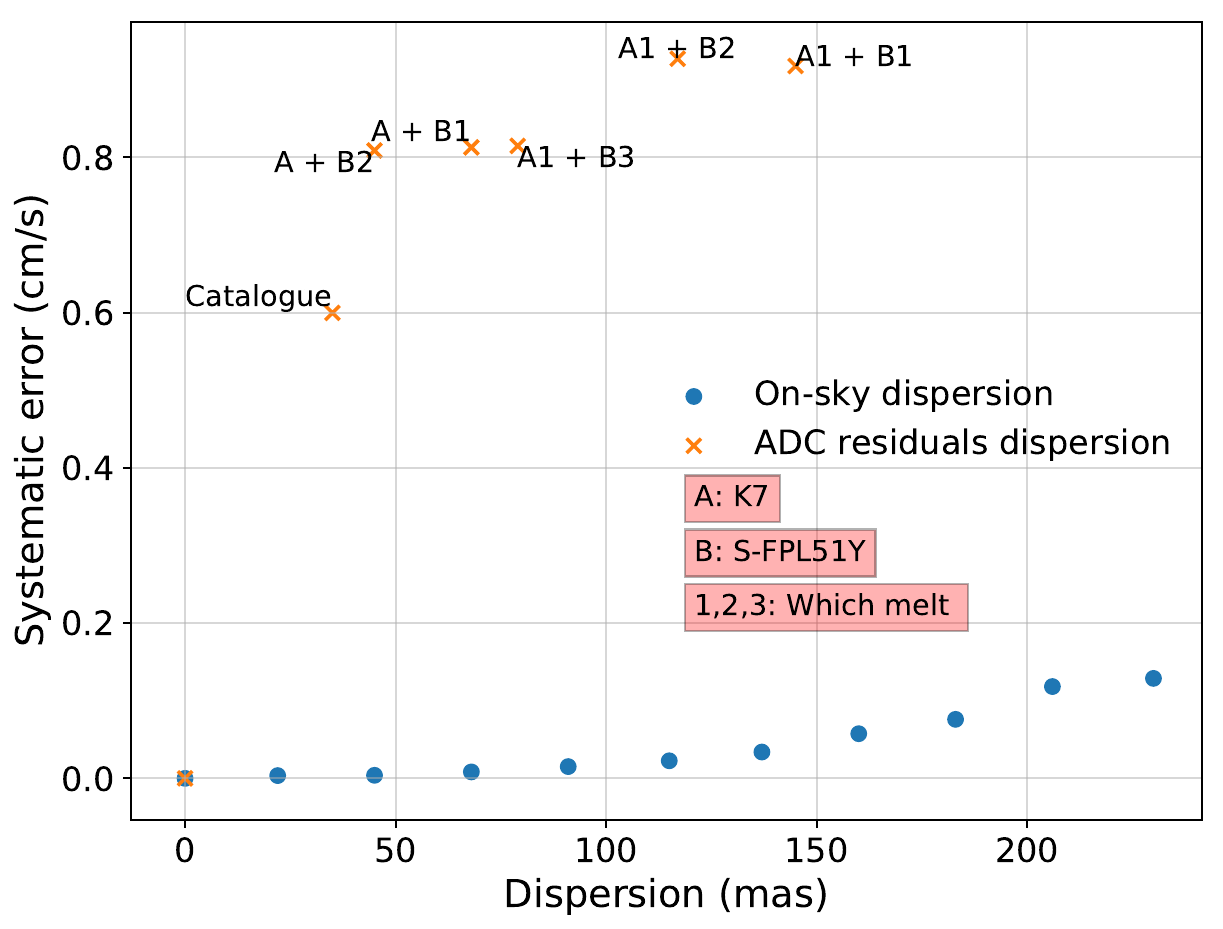}
   \caption{Similar to figure \ref{Fig:systematic_nonoise}, but but adding the different melt data residuals results. The points from Figure \ref{Fig:systematic_nonoise} represent the case where an ADC is used.}
   \label{Fig:systematic_nonoise_melt}
\end{figure}

The results from Figure \ref{Fig:systematic_nonoise}, also represent the case where an ADC is used. Each zenithal angle in that case will represent an amount of dispersion residuals as indicated in Table \ref{Table:disp}. It is clear from Figures \ref{Fig:melt_data_flux_loss} and \ref{Fig:systematic_nonoise_melt}, that ADC residuals up to 100 mas will create an error of $\sim$ 1 cm/s (mainly due to melt data since they produce different residuals shapes) and $\sim$ 1\% on the RV precision and flux losses, respectively. It is also clear from the same figures, that different melt data can return different residuals, different RV variation and different flux losses. It is important, when designing an ADC, to give more attention to the melt data as they can cause a serious problem specially when it comes to flux losses that seems more critical than RV variation.

\section{Conclusions}
\begin{figure}
   \centering
   \includegraphics[width=\hsize]{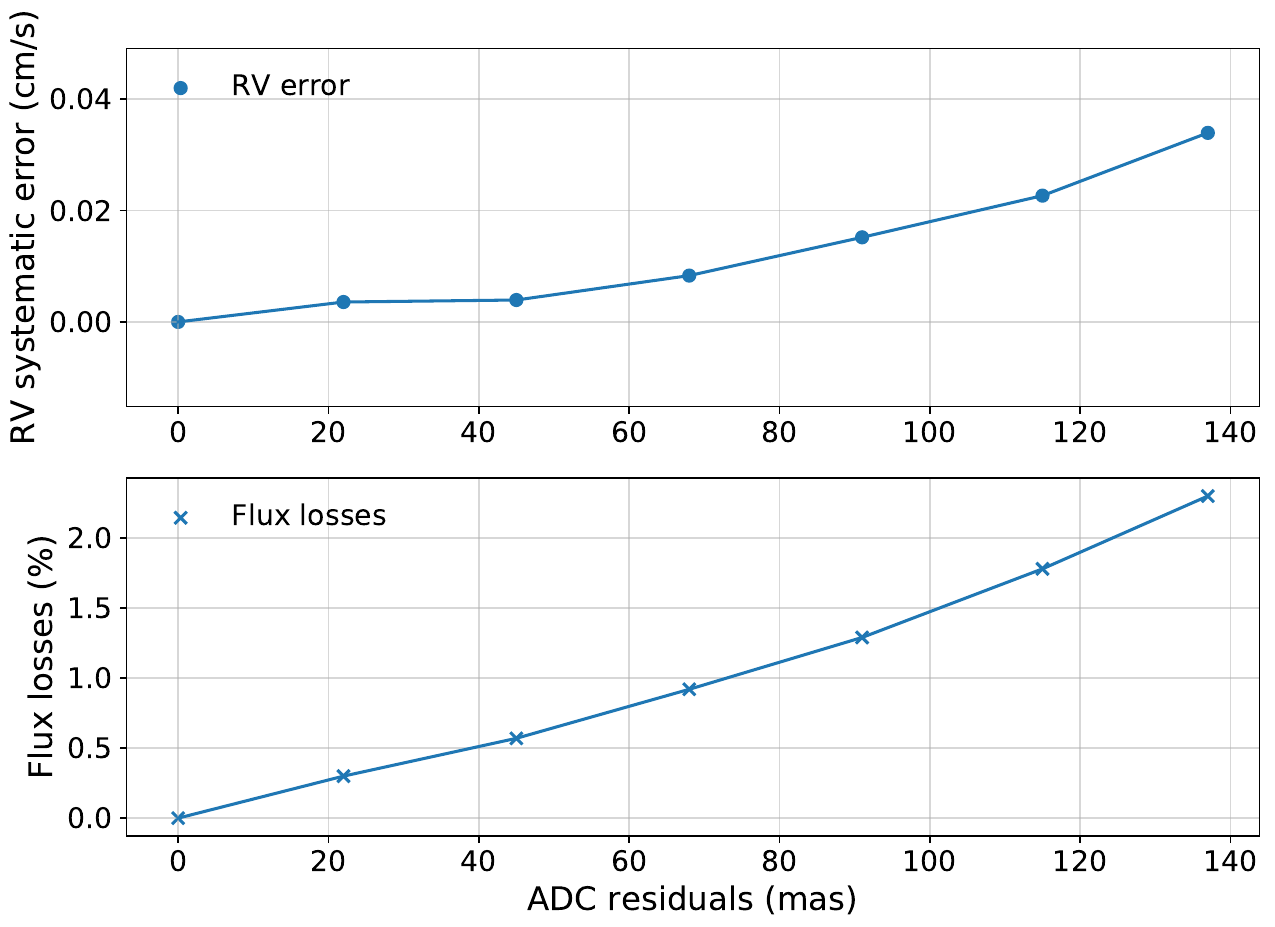}
   \caption{Conclusion plot summarizing the requirements on ADC residuals. Top: expected RV error due to ADC residuals (similar to Figure \ref{Fig:systematic_nonoise}); bottom: expected flux losses at 380 nm due to ADC residuals (similar to Figure \ref{Fig:flux_loss_max}).}
   \label{Fig:conclusion}
\end{figure}

As we showed above, atmospheric dispersion will largely contribute to the errors if not properly corrected by an ADC. The residuals of an ADC will affect the RV precision as well as the flux losses when designing a high-resolution spectrograph. Focusing on the spectral range between 380 nm and 780 nm, the case used in this analysis, ADC residuals of the order of 100 mas are guaranteed not to introduce large errors that could impact a precision of the level of 10 cm/s. In fact, 100 mas PV, will have an impact of $\sim$ 1 cm/s in terms of RV precision. This same 100 mas, will have an impact of $\sim$ 1\% in flux losses. The latter could be the most important factor of the requirements on the ADC specifications. These results are summarized in Figures \ref{Fig:systematic_nonoise_melt}, and \ref{Fig:conclusion}. 

We also noticed that the choice of the guiding camera can play a role and affect specially the flux losses. We presented a study that will allow to define the specifications of an ADC based on the requirements on RV precision and flux losses, taking into account the guiding system, responsible of the fiber-injection, and on the optical optimization based on melt data. This procedure will allow to define the specifications of an ADC in a more accurate way avoiding the (typical) conservative approach that increases (unnecessarily) the complexity and the cost of an ADC.

\section*{Acknowledgements}
The first author is supported by an FCT fellowship (PD/BD/135225/2017), under the FCT PD Program PhD::SPACE (PD/00040/2012). Jorge H. C. Martins is supported in the form of a work contract funded by national funds through Funda\c{c}\~{a}o para a Ci\^{e}ncia e Tecnologia (DL 57/2016/CP1364/CT0007). This work was supported by FCT/MCTES through national funds and by FEDER - Fundo Europeu de Desenvolvimento Regional through COMPETE2020 - Programa Operacional Competitividade e Internacionaliza\c{c}\~{a}o by these grants: UID/FIS/04434/2019; PTDC/FIS-AST/32113/2017 \& POCI-01-0145-FEDER-032113.




\bibliographystyle{mnras}
\bibliography{wehbereferences} 








\bsp	
\label{lastpage}
\end{document}